\documentclass[prd,superscriptaddress]{revtex4}
\usepackage{amstext,amsmath,amssymb,amsfonts}
\usepackage{bm,bbm}
\usepackage{epsfig}
\usepackage{verbatim}

\usepackage[normalem]{ulem}

\def\be{\begin{equation}}
\def\ee{\end{equation}}
\def\ba{\begin{array}}
\def\ea{\end{array}}
\def\bes{\begin{eqnarray}}
\def\ees{\end{eqnarray}}
\def\nn{\nonumber}

\def\6{\langle}
\def\9{\rangle}

\def\pp{\partial}

\def\1{{{\mathbbm 1}}}

\def\f{\frac}

\def\t {\tilde}

\def\cc{{\cal C}}

\def\rr{{\cal R}}

\def\hh{{\cal H}}

\def\com{{\mathbb{C}}}

\def\hH{{\hat{H}}}
\def\hK{{\hat{K}}}
\def\hP{{\hat{P}}}
\def\hR{{\hat{R}}}
\def\hU{{\hat{U}}}
\def\hV{{\hat{V}}}
\def\hY{{\hat{Y}}}
\def\hph{{\hat{\phi}}}

\def\half{\mbox{$\f 1 2$}{}}

\begin{document}
\title{Dynamics and entanglement in spherically symmetric quantum gravity}
\author{Viqar Husain}
\affiliation{Department of Mathematics and Statistics, University of New
Brunswick,\\
  Fredericton, NB  E3B 5A3, Canada  }
\author{Daniel R. Terno}
\affiliation{Centre for Quantum Computer Technology, Faculty of Science,
Macquarie University, \\ Sydney, New South Wales  2109, Australia}

\begin{abstract}
The gravity-scalar field system in spherical symmetry provides a natural setting for
exploring gravitational collapse and its aftermath in quantum gravity.
In a canonical approach, we give constructions of the  Hamiltonian operator, and of
semiclassical states peaked on constraint free data.  Such states provide explicit examples
of physical states.  We also show that matter-gravity entanglement is an inherent feature
of physical states, whether or not there is a black hole.
\end{abstract}


\maketitle

\section{Introduction}

Black hole thermodynamics,  cosmology and more generally unification,
have all provided strong impetus for developing a  theory of quantum gravity.  There appears to be agreement on at least two features such a theory should have regardless of the details of the approach.
These are background (or metric) independence and   fundamental  discreteness.  It
may be that other ideas such as holography arise as a consequence of the first two but this is far from clear at the present level of understanding.

Progress in quantum gravity has often been limited to models where a complete quantization can be performed, such as the work on mini-superspace cosmology in the early seventies \cite{mini-70s}, and its later incarnations in the Hartle-Hawking approach \cite{mini-hh} and  loop quantum cosmology\cite{lqc}. These are all quantum mechanical systems arrived at by the assumption of spatial homogeneity of spacetime.  The extent to which such models can reveal  insights into quantum gravity is not clear \cite{kuchar-ryan} although recent developments on singularity avoidance may be one such feature.  More generally the hope is that such smaller systems will ultimately become absorbed in a full theory of quantum gravity in the same way that the Bohr atom
has in  quantum electrodynamics.

Going beyond mini-superspace models significantly enlarges the problem to a true field theory of a constrained system.
The first step in this direction is to a two-dimensional field theory, where the metric and matter variables depend on time and one spatial coordinate. There are a few models of interest in this category, namely the Gowdy cosmology, cylindrical gravitational waves and the asymptotically flat gravity-scalar field theory.  The quantum theory of the first two models have been studied in the simplified approximation of one local degree of freedom \cite{cylindrical,gowdy}. These are vacuum models so they are  less interesting physically  than the last  one, which has the potential to reveal much about gravitational collapse and black hole formation  in quantum gravity \cite{hw-bh, hw-flat, boj-swid}.  Furthermore this model is  the natural next step beyond just the quantum mechanics of the  Schwarzschild black hole, on which much has been written \cite{geg-kun,kuchar-bh,thiemann-bh,ash-boj-bh,modesto-bh, gp-bh}.

This system has  a true  Hamiltonian that forms a part of its asymptotic Poincar\'{e} symmetries.  The model provides the setting for two important physical scenarios:  numerical studies of gravitational collapse \cite{choptuik,gundlach-rev} and  the so-called information loss paradox, whose origin involves an  assumption on the large scale structure of an evaporating black hole based on semiclassical physics \cite{info-loss}. It is concerned with a collapsing  low-entropy (or even pure-state) matter that forms a black hole, which subsequently evaporates   within a finite time, resulting in a highly-entropic Hawking radiation \cite{h:rad}. If the correlations between the inside and outside of the black hole are not restored during the evaporation process, then the corresponding increase in entropy is interpreted as a lost ``information".  Various arguments have been advanced to justify or dismiss this phenomenon \cite{unibh}, as well as possible consequences of either option \cite{beni}. There is the  possibility  that the existence of a quantum theory of gravity automatically removes the information loss ``paradox",  since a  complete quantization  of the true Hamiltonian would result in the unitary evolution of the combined matter-gravity system \cite{idea}, while the entropy flow between  subsystems can be interpreted with the help of quantum information theory \cite{nc}.

In this paper we build on recent developments  \cite{hw-bh, hw-flat} aimed at obtaining a compete quantum  theory of the asymptotically flat gravity-scalar field theory in spherical symmetry. This earlier
work used a gauge appropriate for Painleve-Gullstrand coordinates (because of their regularity at the horizon), and provided a construction of the Hamiltonian and null expansion operators.  Not addressed
in these works  are the problems of finding physical states and addressing their dynamics.  This is what we study  in this paper.

The outline of this paper is as follows.  We first review the model and its quantization. We then  give a prescription for  constructing semiclassical physical states that are peaked on  classical data, and give a new form of the Hamiltonian operator. (An earlier version of this operator utilized a Dirac-like  trick to write a square root operator.) Using these we show that physical states exhibit matter-gravity entanglement. We argue that this may be a robust model independent  feature of non-locality in quantum gravity

\section{Gravity-scalar field model}

  Our starting point   is the Arnowitt-Deser-Misner (ADM) hamiltonian formulation for general relativity.
 The phase space of the model is defined by prescribing a
form of the gravitational phase space variables $q_{ab}$ and
$\tilde{\pi}^{ab}$, together with fall-off conditions  for these
variables, and for the lapse and shift functions $N$ and $N^a$.  The
  ADM 3+1 action for general relativity minimally coupled to a
massless scalar field is
\be
S = \frac{1}{16\pi G}\int d^3x dt\left[ \tilde{\pi}^{ab}\dot{q}_{ab} +
\t{P}_\phi\dot{\phi}
- N H - N^a C_a\right].
\label{act}
\ee
The pair $(\phi,P_\phi)$ are the scalar field canonical variables, and
\bes
{ H} &=& \frac{1}{\sqrt{q}}\left(\tilde{\pi}^{ab}\tilde{\pi}_{ab}
-\frac{1}{2} \tilde{\pi}^2 \right)
                     -  \sqrt{q}R(q) \nn\\
          & &+8\pi G \left(\frac{1}{\sqrt{q}}\tilde{P}_\phi^2
                       + \sqrt{q}q^{ab}\pp_a\phi\pp_b\phi\right) \simeq 0  \\
{ C}_a &=& D_c\t{\pi}^c_a - 8\pi G\t{P}_\phi\pp_a\phi \simeq0,
\ees
where $\t{\pi}=\t{\pi}^{ab}q_{ab}$.

This action (together with the boundary terms, see e.g. \cite{hw-flat, eric}) is well-defined and determines the fall-off conditions on canonical variables. Below (Sec IIA) we outline the structure that results from imposition of spherical symmetry and  use of the flat-slice partial gauge-fixing. (This is a summary of the work in \cite{hw-flat}.) The reason for utilizing this gauge is that the horizon is not located at a coordinate singularity. In Sec IIB we present a family of initial data that  will illustrate, in a later section, a construction  of semiclassical physical states of the model.

\subsection{Hamiltonian theory in the flat slice formalism}

  The reduction to spherical symmetry utilizes an  auxiliary
flat Euclidean metric $e_{ab}$ and unit radial normal  $s^a= x^a/r$, where  $r^2=e_{ab}x^ax^b$.
The parametrization of the reduced phase space we use is given by the matrices
\bes
q_{ab} &=& \Lambda(r,t)^2\ s_a s_b + \frac{R(r,t)^2}{r^2}\ ( e_{ab} - s_a
s_b),\\
\t{\pi}^{ab} &=& \frac{P_\Lambda(r,t)}{2\Lambda(r,t)}\ s^a s^b + \frac{r^2
P_R(r,t)}{4R(r,t)}\
(e^{ab} - s^a s^b),
\label{reduc}
\ees
As a line element the spatial metric is therefore
\be
ds^2 = \Lambda^2(r,t)  dr^2 + R(r,t)^2 d\Omega^2,
\ee
where the solid angle $d\Omega^2$ arises from the  second term in these expressions.

The Painleve-Gullstrand  (PG) coordinates are those  where equal coordinate time slices are
spatially flat \cite{hw-flat,eric}.  However it is sufficient to use the partial gauge fixing
$\Lambda=1$  to obtain the feature of non-singular coordinates at the horizon, which is
the feature of PG coordinates we desire. This is what is summarized in the remaining part
of this section.

 Substituting (\ref{reduc}) into the ADM $3+1$ action with minimally coupled
scalar field  leads to the reduced action
\bes
S_R &=& \frac{1}{4G}\int dtdr \left(P_R\dot{R} + P_\Lambda\dot{\Lambda} +
P_\phi\dot{\phi}
    \right)\nn\\
     && -\frac{1}{4G}\int dtdr \left( NH + N^r C_r\right) \nn\\
 && -\int dt(N^r \Lambda P_\Lambda)|_{r=\infty},
 \label{red-action}
 \ees
where $N,N^r$ are the lapse and radial shift functions,  and the reduced Hamiltonian and (radial) diffeomorphism  constraints $H$ and $C^r$ are
\bes
H &=& \frac{1}{R^2\Lambda}\left[\frac{1}{8}\ (P_\Lambda \Lambda)^2 -
\frac{1}{4}(P_\Lambda \Lambda)(P_R R)\right]
\nn\\
&&+ \frac{2}{\Lambda^2}\left[ 2RR''\Lambda -2RR'\Lambda' - \Lambda^3 +
\Lambda R'^2 \right] \label{Hsph}\nn\\
&& + \left[\frac{P_\phi^2}{2\Lambda R^2} + \frac{R^2}{2\Lambda}\ \phi'^2
\right]\simeq 0,\\
C_r &=&   P_R R'  -\Lambda P_\Lambda' + P_\phi\phi' \simeq 0.
\label{Csph}
\ees
These   constraints are first class with an algebra that is similar to that for the full theory.

We note that since this reduction to spherically symmetry is unusual in that the usual ADM mass
integral vanishes due to the flat slicing condition. It is therefore important to ensure that the reduced action in functionally differentiable with a consistent set of fall-off conditions  on the phase space variables, and on the lapse and shift functions. This analysis has been carefully done by one of the authors in Ref.  \cite{hw-flat}, where these details are explicitly given.  The asymptotic conditions
on the phase space variables are
\bes
R &=& r +{\cal O}(r^{-1/2-\epsilon}),\nn \\
P_R &=& Ar^{-1/2}/2 + {\cal O}(r^{-1-\epsilon}), \label{R}\\
\Lambda &=& 1 + {\cal O}(r^{-3/2-\epsilon}),\nn \\
P_\Lambda &=& A r^{1/2} + {\cal O}(r^{-\epsilon}) \label{Lam}\\
\phi &=& Br^{-1/2} + {\cal O}(r^{-3/2-\epsilon}),\nn \\
P_\phi &=& Cr^{1/2} + {\cal O}(r^{-\epsilon}),
\label{phi}
\ees
and those on the  lapse and shift functions are
\bes
N^r &=& Ar^{-1/2} + {\cal O}(r^{-1/2-\epsilon}), \nn\\
N &=& 1 + {\cal O}(r^{-\epsilon}),
\label{asympN}
\ees
where $A,B,C$ are constants. The leading order terms are solutions of the constraints, so there
is no logarithmic divergence of the action at this order. The mass formula is the surface term in Eqn. (\ref{red-action});  the mass parameter arises in this integral through the momentum $P_\Lambda$. It is readily verified that if the scalar field is set to zero the Schwarzschild solution results \cite{hw-flat}.

\subsection{Partial gauge fixing}

We next impose the  gauge choice $\Lambda=1$, which corresponds to a step toward flat slice coordinates.   With this   condition imposed, the Hamiltonian constraint is solved (strongly) for the conjugate momentum $P_\Lambda$ as a function of the phase space variables.  This gives
\be
P_\Lambda = P_RR + \sqrt{ (P_RR)^2 - X},
\label{PL}
\ee
where
\be
X =  16R^2 (2RR'' - 1 + R'^2) + 16R^2 H_\phi,
\ee
and
\be
 H_\phi = \frac{P_\phi^2}{2R^2} + \frac{R^2}{2}\ \phi'^2.
\ee
This effectively eliminates the conjugate pair $(\Lambda, P_\Lambda)$ in favour of the remaining
phase space variables.  We must however obtain the consequence of this for the lapse and shift functions in the standard way: the evolution equation for $\Lambda$ \cite{hw-flat} and the requirement that the gauge $\Lambda=1$ be preserved under evolution leads to the equation
\be
N = -\frac{4R^2 (N^r)'}{\sqrt{(P_RR)^2 - X}}.
\label{lapse}
\ee
We note that this relation does not lead to any restriction on the phase space variables, since it is merely a relation between the lapse $N$ and and the shift $N^r$, neither of which is absolutely fixed at this stage.  We also note that on the classical space of solutions, the argument of the square root is positive;
in quantum theory, it is in principle possible that fluctuations may make this negative. This however is an issue that is present in general in the ADM variable approach to quantum gravity because the Hamiltonian constraint contains a factor of $\sqrt{q}$, which may fluctuate such that the argument is negative. The issue hinges on how the corresponding operator is defined. We discuss this below   where we set up the Dirac quantization problem.

The gauge $\Lambda=1$ reduced the radial diffeomorphism constraint to
\be
C_r = -P_\Lambda' + P_R R' + P_\phi \phi' \simeq 0,
\ee
with $P_\Lambda$ given by (\ref{PL}) above. We note that using this constraint  the square root in the latter equation
can be written as
\be
 \sqrt{ (P_RR)^2 - X} = \int_0^r \left(    P_R R' + P_\phi \phi'  \right) -P_RR
 \label{root-expr}
\ee

To summarize this section,  the partially gauged fixed theory is prescribed by the phase space variables
$(\phi,P_\phi)$ and $(R,P_R)$ and the reduced Hamiltonian
\bes
H_{R}^G &=& \int_0^\infty \left[(N^r)'P_\Lambda + N^r(P_RR' + P_\phi \phi')\right]dr \nn\\
&=&\int_0^\infty (N^r)'\left( R P_R + \sqrt{(P_RR)^2 - X}\right) dr\nn\\
  && + \int_0^\infty N^r(P_RR' + P_\phi \phi')\ dr,
\label{Hred}
\ees
where the surface term in the reduced action (\ref{red-action}) has been written as bulk term and combined with the remaining radial diffeomorphism constraint.  Together with the asymptotic conditions given above, the variational principle is well defined. This is the reduced system we study in the rest of the paper.

Finally we note that we have not yet fully fixed the gauge to the flat slice case
since we have not imposed the condition $R=r$. We prefer to retain $R$ as a dynamical variable
for developing the quantum theory, since full gauge fixing  leads to a reduced Hamiltonian  for the scalar field degrees of freedom that  is non-local \cite{hw-flat}, and so poses a problem for quantization.

\subsection{Classical data}

Finding classical initial data sets  is necessary both in numerical calculations and for the construction of semi-classical physical states. As one of our aims in this paper is to present a concrete realization
of the latter, it is useful to see what form such data takes before we proceed to a construction of such states. The data  should be an asymptotically flat  solution of the remaining  constraints  prescribed by functions $\phi(r),P_\phi(r)$ and $R(r)$ and $P_R(r)$.    The family of data that we  exhibit below depends on two parameters. It is regular at the coordinate origin and may  have trapped surfaces, depending on the parameter values.

We shall see, after describing the quantum theory, that semiclassical states corresponding to
data with trapped surfaces such as this, describe a quantum black hole state. This occurs because  states peaked on classical constraint free data satisfy the condition that the expectation value of the constraint operator vanishes  to leading order in a well defined expansion (see Section V below.)

 With the partial gauge fixing $\Lambda =1$ already imposed as described in Sec. II.B, we fix gauge
 fully by imposing $R=r$. This simplifies   the constraint to
\be
C_r=-P_\Lambda' + P_R + P_\phi \phi' =0, \label{difco}
\ee
with
\be
P_\Lambda = P_Rr +\sqrt{(P_Rr)^2 -X}, \label{pl}
\ee
where $X$ now reduces   to
\be
X\equiv 16R^2(2RR'' - 1+ R'^2) + 16R^2 H_\phi= 16r^2 H_\phi.
\ee
To find some explicit solutions of interest, we make the ansatz $\phi=0$. The constraint then
becomes a relation between $P_R$ and $P_\phi$, so one of these may be chosen freely.

To obtain an explicit class of solutions, let us set
\be
16H_\phi=h^2(r)P^2_R(r), \label{Hphi-ansatz}
\ee
for some function $h$. Then (\ref{pl}) becomes
\be
P_\Lambda=P_Rr(1+\sqrt{1-h^2})\equiv P_Rr(1+g),
\ee
and the auxiliary function $h=\sqrt{1-g^2}$ satisfies
\be
0\leq h^2\leq 1,\qquad \lim_{r\rightarrow\infty}h^2=0.
\ee
The constraint is now
\be
P'_Rr+(P_Rrg)'=0,
\ee
which can be rewritten as
\be
(P_Rr)'-P_R+(P_Rrg)'=0.
\ee
Finally, this may be put into an integrated form by setting
\be
P_R = \Pi_R'
\ee
to give
\be
\Pi_R=c\exp\left(\int^r\!\!\!\frac{dx}{(1+g(x))x}\right),\qquad.
\ee
Thus, given suitable functions $g$, we can find $\Pi_R$ and hence $P_R$, followed
by  $P_\phi$ from Eqn. (\ref{Hphi-ansatz}).

We note that the energy density can be written as
\be
16H_\phi=2\frac{\Pi_RP_R}{r}-\frac{\Pi_R^2}{r^2}.\label{hphi}
\ee
The asymptotic fall-off conditions (above) require $P_R\sim 1/\sqrt{r}$,
which implies $\Pi_R\sim 2P_Rr$ and $g\rightarrow 1$. If one wishes
to have a regular data at $r=0$, then the expansion $\Pi_R\sim
r^\alpha$   leads to
\be
0\leq  \frac{1}{\alpha}-1\leq 1,
\ee
which bounds the power to
\be
1/2\leq \alpha \leq 1,
\ee
while the regularity of $H_\phi$ imposes $\alpha=1$, so for $r\sim
0$ we get $\Pi_R\sim c r$
\begin{figure}[floatfix]
    \begin{minipage}{\columnwidth}
    \begin{center}
        \resizebox{0.5\columnwidth}{!}{\includegraphics{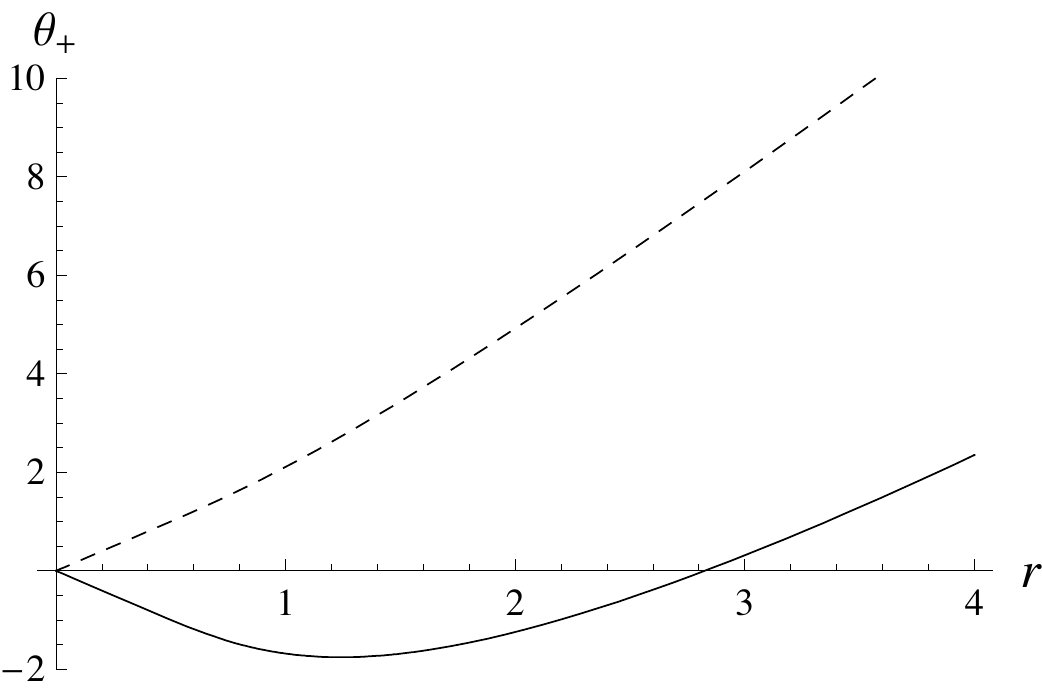}}
    \end{center}
    \end{minipage}
    \caption{\small{Expansion of the outgoing null geodesic with (solid line) and
without (dashed line) trapped surfaces on the initial hypersurface.
The solid line corresponds to $a=1, C=2$ and the dashed line to
$a=1, C=6)$.
} }\label{trapplot}\vspace{-5mm}
\end{figure}

To obtain an explicit asymptotically flat solution let us consider the choice
\be
1+g=(1-\half\exp(-a/x^2))^{-1},
\ee
which results in
\be
\Pi_R=cr\exp\left[-\frac{1}{4} {E}_1\!\left(\frac{a}{r^2}\right)\right],
\ee
where the integral exponential function is defined as
\be
E_1(z)=\int_z^\infty\!\frac{e^{-t}}{t}dt.
\ee
The asymptotic behavior of $\Pi_R$ is as desired,
\be
\Pi_R\approx c\,^4\!\!\!\sqrt{a}e^{\gamma/4}\sqrt{r}, \qquad
r\rightarrow\infty,
\ee
where Euler's gamma is $\gamma=0.57\ldots$, and
\be
\Pi_R\approx cr, \qquad r\rightarrow 0
\ee
From the asymptotic fall-off condition
\be
P_R=Ar^{-1/2}/2+{\cal{O}}(r^{-1-\epsilon}),
\ee
where $A$ is related to the ADM mass as $A=4\sqrt{2M}$
\cite{hw-flat}, we see that $c\,^4\!\!\!\sqrt{a}e^{\gamma/4}=A$.

 The next step in the investigation of the classical problem is to
check for the presence of apparent horizons. In the  gauge we
adopted the expansions of outgoing and ingoing null geodesics are given
as  \cite{hw-bh}
\be
\theta_{\pm}=\pm 4RR'- P_\Lambda,
\ee
that reduces for the initial data surface to
\be
\theta_{\pm}=\pm 4r- \Pi_R
\ee
By varying parameters of $\Pi_R$ it is possible to generate the
regular initial data with or without trapped surfaces, as shown on
Fig.~\ref{trapplot}.

\section{Quantization}

In this section we describe an approach to constructing the quantum theory of the system described above. We will be working with the partially gauge fixed theory described in Sec. II.B, where we have
the pair of phase space variables $(R,P_R)$ and $(\phi,P_\phi)$, and the remaining radial
diffeomorphism constraint. The quantization we use is a type of polymer quantization where
field configuration and translation operators   are  quantized, but there is no direct quantization
of field momenta.  Using the basic operators, we shall see that there are well defined  prescriptions
for writing the constraint and the Hamiltonian operators.

\subsection{Hilbert space and basic operators}
To describe the polymer quantization \cite{hw_quant, quant} (which may be viewed as the ``dual''   to that used in loop quantum gravity), we begin with the basic variables
\be
R_f = \int_0^\infty dr  f(r) R(r), \ \ \ \ \ \ \ \  U_\epsilon(P_R)= \exp\left(i\epsilon P_R\right),
\label{basic-var}
\ee
where $f(r)$ is a smearing function and $\epsilon$ is a  real dimensional constant.  These satisfy the canonical Poisson bracket
\be
\{ R_f , U_\epsilon(P_R(r)) \}  = i \epsilon f(r) U_\lambda(P_R(r)).
\label{basicpb}
\ee
(In defining these observable we have at our disposal a constant  density that ensures the correct
density weights in the integral for $R_f$ and  in the exponent of $U_\epsilon$.  This density is the
phase space variable $\Lambda$ which was set to unity when we performed the partial gauge
fixing $\Lambda=1$. )

We use  similar definitions for the variables made from $\phi$ and $P_\phi$, ie.
 \be
\phi_f = \int_0^\infty dr  f(r) \phi(r), \ \ \ \ \ \ \ \  V_\kappa(P_\phi)= \exp\left(i\kappa P_\phi\right). \nn
\label{basic-var-matter}
\ee
The Poisson bracket of these matter variables mirrors that of the metric variables in
(\ref{basicpb}).

The parameters $\epsilon$ and $\kappa$ have the physical dimensions necessary to make the exponents in the corresponding $U$ variables dimensionless.  This scale will carry over to quantization. Since we are dealing with a quantization of gravity and matter, it is natural to suppose that that both parameters are related to the Planck scale. In the following we will keep them separate.

The Poisson bracket of the metric variables is  realized as an operator relation on a Hilbert space spanned by the  basis states
\be
| a_1,a_2,\cdots ,a_n\rangle \label{basis}
\ee
where the real numbers $a_i$ represent values of the configuration variable $R$ at the radial points $r_i$.  The points $r_i$  provide a set where the phase space variables are sampled. The selection of space points $r_i$  may be thought of as a lattice (or graph in the language used in loop quantum gravity). Since there is a basis state of this type for every countable set of points, the Hilbert space  is non-separable. In the following it will be convenient to work with a fixed and uniformly spaced  set of points, although this is not necessary.

The inner product  is
\be
\langle a_1',a_2',\cdots ,a_n'| a_1,a_2,\cdots ,a_n\rangle= \delta_{a_1',a_1}\cdots \delta_{a_n',a_n}
\ee
if two states are associated with the same  lattice points; if not the inner product is zero.  This inner product is background independent in the same way  as for example the inner product for the Ising model; the difference is that for the latter there is a finite dimensional space of  spins at each lattice point.

The configuration and translation operators are defined by the following expressions:
\be
\hat{R}_f| a_1,a_2,\cdots a_n\rangle:= \sum_i a_i f(r_i)| a_1,a_2,\cdots a_n\rangle,
\ee
\be
\widehat{U}_\epsilon (P_R(r_k))| a_1,a_2,\cdots a_n\rangle := | a_1,a_2,\cdots , a_k- \epsilon, \cdots a_n\rangle.
\label{trans-op}
\ee
It is readily verified that the commutator of these operators as defined provides  a faithful realization of the corresponding Poisson bracket.  Since the $U$ operators are realized here as ladder operators for the field excitations, the parameter $\epsilon$ represents the discreteness scale in field configuration space.

This representation is one in which the momentum operator does not exist. There is however
an alternative $\epsilon$ dependent definition of momentum using the translation operators
(\ref{trans-op}), given by
\be
\hat{P}_R^\epsilon(r_k) := \frac{1}{2i\epsilon}\ \left(\hat{U}_\epsilon(P_R(r_k)) - \hat{U}_\epsilon^\dagger (P_R(r_k)) \right),
\label{mom}
\ee
which will be used in the definition of  the Hamiltonian operator.

The representation for the matter variables is similar to that defined above for the metric variables.
We write the basis states of the matters sector as
\be
| b_1,b_2,\cdots ,b_n\rangle \label{matter-basis}\nn
\ee
With the inclusion of matter the kinematical Hilbert space is the  tensor product of geometry
and matter Hilbert spaces, with basis
\be
|\underbrace{a_1,\ldots, a_N}_{gravity};\underbrace{b_1,\ldots,
b_N}_{matter}\9, \label{full-basis}
\ee

With the quantization of the basic variables in hand, we now turn to defining the composite operators
necessary to write the constraints  and the Hamiltonian. To this we first need to define a localized
field operator from $R_f$ and $\phi_f$. A localized field  may be defined by taking for example $f(r)$ in eqn. (\ref{basic-var}) to be a Gaussian
\be
G(r,r_k,\sigma) = e^{ - (r-r_k)^2/\sigma^2},
\ee
(or a smooth function of bounded support) which is sharply peaked  at a radial point $r_k$. For illustration we will work with the Gaussian with the understanding that its width $\sigma$ is such
that the function is effectively zero at all lattice points except where it is peaked. This is easiest to see
with a uniform lattice such that $\sigma\ll 1$ in Planck units. With this in mind we write
\be
R_{G(r_k)}\equiv R_k,
\ee
with a similar expressions for the scalar field $\phi$. We also  have the following
action of the basic operators (where we have included explicitly the Planck length):
\be
\hR_k|{a_1,\ldots, a_N};{b_1,\ldots,
b_N}\9=2l_P^2a_k|{a_1,\ldots, a_N};{b_1,\ldots, b_N}\9,
\ee
 \be
\hph_k|{a_1,\ldots, a_N};{b_1,\ldots, b_N}\9=2l_P^2b_k|{a_1,\ldots, a_N};{b_1,\ldots,
b_N}\9.
\ee
The field translations operators $\hU_k(\epsilon)\equiv\widehat{e^{i\epsilon
P_{Rk}}}$ and $\hV_k(\kappa)\equiv\widehat{e^{i\kappa P_{\phi k}}}$
act as
\be
\hU_k(\epsilon)|{a_1,\ldots, a_N};{b_1,\ldots,b_N}\9=|{a_1,\ldots,a_k-\epsilon,\ldots a_N};{b_1,\ldots,
b_N}\9
\ee
and
\be
\hV_k(\kappa)|{a_1,\ldots, a_N};{b_1,\ldots,b_N}\9=|{a_1,\ldots
a_N};{b_1,\ldots,b_k-\kappa,\ldots, b_N}\9
\ee
Thus,  with the choice of Gaussian smearing functions sharply peaked at the points $r_k$ in the
operators $\hat{R}_f$ and $\hat{\phi}_f$, the basic commutators are
\be
[\hR_k,\hU_l(\epsilon)]=-2l_P^2\epsilon\delta_{kl}\hU_l(\epsilon),
\qquad [\hph_k,\hV_l(\kappa)]=-2l_P^2\kappa\delta_{kl}\hV_l(\kappa)
\ee

Since $\hR_k$ operators have zero in the spectrum (ie. there are states such as $|a_1 \cdots, a_k=0,\cdots a_n\rangle$
that have zero eigenvalue), there is no  inverse operator and an indirect definition is required. This is achieved by
Poisson bracket identities such as
\be
\{\sqrt{|R_k|},U_l(\epsilon)\}=\f{i}{2\sqrt{|R_k|}}\epsilon
U_l(\epsilon)\delta_{kl},
\ee
first noted by Thiemann.  The operator
\be
\hK_k\equiv\widehat{\f{1}{R_k}}\equiv\left(\f{2}{2l_P^2\epsilon^2}\hU_k(-\epsilon)\left[\sqrt{|\hR_k|},\hU_k(\epsilon)\right]\right)^2
\ee
depends on the parameter $\epsilon$ through its dependence on the translation operators $\hU_k\epsilon$.
 Its action on the basis states is
 \be
\widehat{\f{1}{R_k}}|{a_1,\ldots,
a_N};{b_1,\ldots,b_N}\9=\f{1}{2l_P^2\epsilon^2}(\sqrt{|a_k-\epsilon|}-\sqrt{|a_k|})^2|{a_1,\ldots,
a_N};{b_1,\ldots,b_N}\9.
 \ee
A symmetric version of this operator is similarly defined.

Let us now turn to realizing operators corresponding to  spatial derivatives of field variables such as
$\phi'$ and $R'$.  To do this in a controlled manner it is easier to first restrict the choices of points $r_i$
so that they define a uniformly spaced lattice with spacing $\lambda$. This is the first place where we explicitly introduce a radial lattice. It also has the effect of selecting a subspace of the full Hilbert space we defined above, by working only with a fixed set of radial points.   We can now follow what is done in numerical methods, ie.
 \be
f'(r_k)\rightarrow \f{f_{k+1}-f_{k}}{l_P\lambda},
\ee
and
\be
f''(r_k)\rightarrow\f{ f_{k+1}-2f_k+f_{k-1}}{2l_P^2\lambda^2},
\ee
for any function $f_k$. This first of these suggests the operator
\be
 \widehat{R'}_k =  \f {\hat{\phi}_{k+1} - \hat{\phi}_k }{l_P \lambda},
\ee
with similar expressions for other operators. Such choices are of course not canonical,
and may be viewed as an additional freedom in realizing  a quantum theory.

Lastly the lattice local obsevable momentum operators  are
\be
\hP_{Rk}\equiv\f{l_P}{2i\epsilon}(\hU_k(\epsilon)-\hU_k^\dag(\epsilon)),
\qquad
\hP_{Rk}^2\equiv\f{l_P^2}{\epsilon^2}(2-\hU_k(\epsilon)-\hU_k^\dag(\epsilon)),
\ee
\be
\hP_{\phi k}\equiv\f{l_P}{2i\kappa}(\hV_k(\kappa)-\hV_k^\dag(\kappa)),
\qquad
\hP_{\phi k}^2\equiv\f{l_P^2}{\kappa^2}(2-\hV_k(\kappa)-\hV_k^\dag(\kappa)),
\ee
with action on basis states given by
\be
\hP_{\phi
k}|a;b\9=\f{l_P}{2i\kappa}(|a;b_1,\ldots,b_k-\kappa,\ldots,b_N\9-|a;b_1,\ldots,b_k+\kappa,\ldots,b_N\9),
\ee
and
\be
\hP_{\phi k}^2|a;b\9=\f{l_P^2}{\kappa^2}(2|a;b\9-|a;b_1,\ldots,b_k-\kappa,\ldots,b_N\9-|a;b_1,\ldots,b_k-\kappa,\ldots,b_N\9)
\ee
In constructing more complicated operators the question of the operator ordering is important. One option is a symmetric ordering,
\be
AB\rightarrow
\widehat{AB}\equiv(\hat{A}\hat{B}+\hat{B}\hat{A})/2 \label{qcon}.
\ee
Another possibility is an order at which, e.g., $\hR$ is to the right of $\hP_R$. We will see in the following that it has an advantage of annihilating the state of zero gravitational excitations.

\subsection{Constraint   operator}

Our goal in this section to use the above definitions of basic operators to give a prescription for the
reduced Hamiltonian (\ref{Hred}).  The main issue is the definition of the square root $\sqrt{Y}$ in this Hamiltonian,
where
\be
Y=(P_RR)^2 -\rr- 16R^2 H_\phi, \qquad \rr=16R^2(2RR'' - 1+ R'^2).
\ee
An operator representing this expression may be defined using a Dirac-like trick by suitably
extending the kinematical Hilbert space \cite{hw_quant}. However there is an alternative way to write the constraint
(\ref{difco})  such that the square root problem is bypassed. Substituting for $P_\Lambda'$ from
Eqn. (\ref{PL})  into the diffeomorphism constraint gives
\be
C_r=-P'_RR-\frac{Y'}{2\sqrt{Y}}+P_\phi\phi'\simeq 0
\ee
If supplemented with the requirement $Y>0$ (which is the case for classical solutions), this is equivalent to
\be
-\half Y'+(P_\phi\phi'-P'_RR)\sqrt{Y}\simeq 0.
\ee
This implies the  constraint
\be
\cc=Y'^2/4-(P_\phi\phi'-P'_RR)^2Y\simeq 0
\label{norootdiff}
\ee
In this form it does not contain a square root and it is now straightforward to construct the corresponding
operator using the basic ones defined above. This path for constructing the constraint operator is one of the  results of this
paper, and provides an alternative to the Dirac-like method used earlier \cite{hw_quant} where a square root operator is
constructed.

Before giving a construction of an operator analog of  (\ref{norootdiff}) , we note however that this constraint may have quantum solutions,
which are not solutions of the original constraint. There is however an obvious check using Eqn. (\ref{root-expr}): after obtaining
a solution, we must check that the expectation value of the operator analog of
\be
-\half Y' - (P_\phi \phi' - P_RR') \int_0^r (P_RR' + P_\phi\phi') dr' \label{expeval}
\ee
in the proposed solution does not vanish. As we discuss below it is possible to construct the corresponding operator.

Let us first focus on the local operator $\hY_k$.
 Its ingredients include
\be
\hR_k^2|a,b\9=4l_P^4a_k^2|a;b\9, \qquad
\hK_k^2|a;b\9=\f{1}{4l_P^4\epsilon^4 }\left(\sqrt{|a_k-\epsilon|}-\sqrt{|a_k|}\right)^4|a;b\9,
\ee
and
\be
(\hph_k')^2|a;b\9=\f{4l_P^2}{\lambda^2}(b_{k+1}-b_k)^2|a;b\9,
\ee
and since all factors of the field Hamiltonian (density) commute,
\be
\hH_{\phi k}=\half \hP_{\phi k}^2 \hK^2_k+\half \hR_k^2(\hph_k')^2.
\ee
To complete the construction of $\hY$ one needs also
\be
(\hR_k')^2|a;b\9=\f{4l_P^2}{\lambda^2}(a_{k+1}-a_k)^2|a;b\9,
\ee
and
\be
\hR_k''|a;b\9=\f{1}{\lambda^2}(a_{k+1}-2a_k+a_{k-1})|a;b\9.
\ee
The last term is
\be
P_R^2R^2\rightarrow \half(\hP_R^2\hR^2+\hR^2\hP_R^2),
\ee
which acts as
\be
\widehat{(P_R^2R^2)}_k|a;b\9=\f{4l_P^6}{\epsilon^2}(2a_k^2|a;b\9-[(a_k-\epsilon)^2+a_k^2]|a_1,\ldots,a_k-\epsilon,\ldots,a_N;b\9
-[(a_k+\epsilon)^2+a_k^2]|a_1,\ldots,a_k+\epsilon,\ldots,a_N;b\9)
\ee
Putting these pieces together we get
\be
\hY_k=\widehat{(P_R^2R^2)}_k-16\hR^2_k(2\hR_k\hR_k'' - 1+ \hR_k'^2) - 16\hR_k^2
\hH_{\phi k}
\ee
To complete the constraint operator $\hat{C}_k$ one also needs the commuting operators
$(\widehat{P_\phi\phi'})_k$, $(\widehat{ P_R'R})_k$. Both are
obtained by, e. g., applying the symmetric quantization condition
Eq.~(\ref{qcon}) to the elementary operators of the previous
section.
\be
(\widehat{P_\phi
\phi'})_k|a;b\9=\f{l_P^2}{i\kappa\lambda}\left.\frac{}{}\!\!\right((b_{k+1}-b_k+\half\kappa)|a;\ldots,b_k-\kappa,\ldots\9
-(b_{k+1}-b_k-\half\kappa)|a;\ldots,b_k+\kappa,\ldots\9\left.\frac{}{}\!\!\right)
\ee
and
\bes
(\widehat{P_R'R})_k|a;b\9 & = & \frac{l_P^2}{i\epsilon\lambda} \left.\frac{}{}\!\!\right(a_k|\ldots,a_{k+1}-\epsilon,\ldots;b\9-(a_k-\half\epsilon)|\ldots,a_k-\epsilon,\ldots;b\9 \nn \\
& & \left.\frac{}{}-a_k|\ldots,a_{k+1}+\epsilon,\ldots;b\9 + (a_k+\half\epsilon)|\ldots,a_k+\epsilon,\ldots;b\9\right)
\ees

\subsection{Hamiltonian operator}

The asymptotic Hamiltonian is the surface term in (\ref{red-action}) for which we need to define an operator for $P_\Lambda$.  A direct quantization of the Hamiltonian density in Eqn. (\ref{Hred}) would require
us to define the operator
\be
\hat{H}^{\mathrm{phys}}_k={N^r}_k'(\widehat{P_RR})_k+|\hat{Y}_k|^{1/2}+N^r_k\left((\widehat{P_R'R})_k+(\widehat{P_\phi
\phi'})_k\right),
\ee
 which has the square root term  $|\hat{Y}_k|^{1/2}$, just as for the constraint. Since
$\hat{Y}_k$ is not diagonal in the basis we are using, this operator is not easy to define unless we go to a different basis.  However, from the constraint $C_r$ we see that classically on the constraint surface we have
\be
P_\Lambda(r)  = \int ^r_0 dr' (P_\phi \phi' + P_RR').
\ee
This suggests that for physical states it is possible to compute the energy by finding an operator analog of the
r.h.s. of this equation. Since the quantization we are using utilizes a radial lattice we can write the integral as a discrete
sum over the lattice points  $r_k$. It is therefore reasonable to suggest the definition
\be
\hat{P}_{\Lambda k} =  \lambda l_P\sum_{i=1}^k  \left[ (\widehat{P_\phi \phi'})_i + (\widehat{P_R R'}_i )\right]
\ee

This operator is useful with the type of semiclassical states we define below in Sec. V. In such states
a computation of the expectation value $\langle \psi | \hat{P}_{\Lambda k}| \psi \rangle$ is possible.  The energy of the quantum spacetime would be this expression evaluated at the farthest lattice point, ie. we would take the limit $k\rightarrow \infty$ after computing the expectation value. This would a possible analog of the classical definition where the energy
\be
E= \lim_{r\to \infty}\ P_\Lambda(r)N^r(r).
\ee
 We note that the asymptotic falloff of $N^r$ is determined by the
classical requirement of functional differentiability \cite{hw-flat}, and this behavior of $N^r$ carries over to the quantum theory, since $N^r$ is not a phase space variable.

\section{Physical states and entanglement}

Initial states of the gravity-matter system should satisfy the quantum constraint
\be
\hat{\cc}|\psi\9=0,
\ee
which is supplemented by the requirement that $|\psi\9$ belong to the positive part of the spectrum of $|\hat{Y}|$. It remains to be seen whether any of the operator ordering choices allows for this to be satisfied on a sufficiently large set of states, or if another  realization of the constraint is necessary to accomplish this.  Nevertheless, it is already possible to make a few remarks. We give
here two observations, on gravity-matter entanglement and  on the information loss problem.

Firstly,   in the case of pure gravity $(\phi=P_\phi=0)$, the ordering that puts the $\hR$ operator to the right results in  the operator form  of the constraint  (\ref{norootdiff}) gives
\be
\hat{\cc}|0\9=\left(\hat{Y}'^2/4-\hat{P}'^2_R\hat{R}^2\hat{Y}\right)|0\9=0,
\ee
where $|0\9$ stands for the state with no excitations, ie. all $a_i=b_i=0$. This may be viewed as a ``degenerate metric vacuum'' because
it is the eigenstate of the field operator $\hat{R}_k$   with zero eigenvalue at all points $r_k$.

Secondly, the presence of the matter-gravity terms in the constraint turns it into entangling operator (see eg. \cite{nc}). This can be seen by considering the constraint in  the form of Eq.~(\ref{norootdiff}). It splits
cleanly into gravity  part, and a separate gravity--matter interaction
\be
\hat{\cc}=\hat{\cc}_G\otimes\1_M+\sum_\alpha\hat{\cc}^G_\alpha\otimes\hat{\cc}^M_\alpha. \label{constrsplit}
\ee
The latter term contains monomials that involve gravity and matter operators. Consider their action on the basis states of  the kinematical Hilbert space $\hh=\hh_G\otimes\hh_M$. The term $\hat{P}_{\phi k} \hat{P}_{R k}$ serves as an example:
\be
\hat{P}_\phi^k \hat{P}_R^k |a\9\otimes|b\9\propto \left(|a_1 \cdots a_k-\epsilon \cdots a_N\9-|a_1 \cdots a_k\9-\epsilon \cdots a_N\9\right)\otimes\left(|b_1\cdots b_k-\kappa \cdots b_M\rangle-|b_1\cdots b_k+\kappa \cdots b_M\rangle\right),
\ee
 which is a direct product of entangled gravity and matter modes. However, adding different terms in the constraint results in a superposition of such direct product states, and the constraint operator cannot be written as $\hat{\cc}=\hat{\cc}_G\otimes\hat{\cc}_M$. Some states of the form $|\psi\9_G\otimes|\varphi\9_M$ may be transformed into a direct product form,
 \be
 \hat{\cc}\left(|\psi\9_G|\varphi\9_M\right)=|\psi'\9_G|\varphi'\9_M,
 \ee
 provided that they satisfy an additional constraint
  \be
 \sum_\alpha\hat{\cc}^G_\alpha\otimes\hat{\cc}^M_\alpha\left(|\psi\9_G|\varphi\9_M\right)\propto |\psi'\9_G|\varphi\9_M.
 \ee
As a result, even if such states exist, they form a lower-dimensional set then the set of all physical states, and by themselves do not form a vector space: their linear combinations are by definition entangled. Since we expect the semiclassical states to be  suitably defined coherent states, we see that these configurations are geometry-matter entangled. This result of course  holds regardless of weather a state corresponds to a classical black hole. The  entanglement   prevents the direct product gravity$-$matter decomposition of a generic physical state.

\section{Semiclassical States}

Semiclassical states that are peaked at a given classical configurations may be defined  for a field theory
just as for a quantum mechanical systems. For a constrained system such as the one we are considering, it is possible to obtain states that are peaked on classical constraint free data. These may be viewed as  approximate physical states in a precise sense.

For a constraint $\hat{C}$ in a field theory, the steps we propose are as follows: (i) begin with an explicit classical solution of the constraints, such as that  given in Sec. II.B, (ii) select sampling points $r_i$ (or a lattice) , and the corresponding solution values  $R(r_i)$,  $\phi(r_i)$ etc. , (iii) construct a Gaussian state peaked at each field value $R(r_i)$ and $\phi(r_i)$, and finally (iv) take the product of the Gaussians, one at each point $r_i$ to give the field theory semiclassical state. This is what we describe in detail below.

Reasonable requirements for an approximate semiclassical state are that
\be
\langle \hat{C}(r_k \rangle = 0, \ \ \ \ \ \ \ \  \langle \hat{C}(r_k)^2 \rangle = 0,
\ee
for each sample point $r_k$. We give here a generalization of the  semiclassical states for Friedmann-Robertson-Walker cosmology  given in  \cite{hw_cosm}, and show how this can be utilized  for the present model.

Let us consider first a single lattice point $r_k$ and the basis states at this point defined by $| a_k\rangle \equiv| m\epsilon \rangle_k$, where $m$ in an integer. Let us consider the state at $r_k$ defined by   the linear combination
\be
 |P_R^0,R^0\rangle^{t,\epsilon}_{r_k} =
\frac{1}{C}\sum_{m=-\infty}^\infty e^{-\frac{t}{2}(\epsilon m)^2}
e^{m\epsilon R^0} e^{im\epsilon P_R^0 }|m\epsilon \rangle_{r_k}.
 \label{semiclass}
\ee
This is a Gaussian  state of width $t$ (measured in Planck units),
 where the  (real) parameters $P_R^0$ and $R^0$ are the
field values corresponding to a classical configuration at the point $r_k$.
The width $t$ is a measure of how strongly peaked the state is on a given classical configuration,
ie. $t \ll 1$ means that fluctuations around $P^0,R^0$ are small.
We shall see in the following that  this state has a number of desirable properties,
and because of this, it allows a construction of approximate physical states of
the theory we are considering. (There may be other non-Gaussian states with such properties
but our purpose here is to point out that this one is useful.)

The normalization condition for the state gives an expression for the
constant $C$.  It is the  convergent sum
\be
C^2=\sum_{m=-\infty}^\infty e^{-t\epsilon^2m^2} e^{2R^0
\lambda m}. \label{norm}
\ee
Calculation with this  state gives the  expectation value \cite{hw_cosm}
\be
\langle \hat{U}_\epsilon \rangle =  e^{i\epsilon P_R^0(r_k)}\
e^{-t\epsilon^2/4} \ K(\epsilon,t, R^0). \label{expecU}
\ee
where
\be
K(\epsilon,t,R^0)=\left(\frac{  1+ 2\sum_{m\ne 0}\cos\left[
\frac{2\pi mR^0}{\epsilon t}\left(1+
\frac{t\epsilon}{2R^0}\right) \right] e^{-\pi^2m^2/t\epsilon^2}
}{  1+ 2\sum_{m\ne 0}\cos\left[ \frac{2\pi mR^0}{\epsilon t}
\right] e^{-\pi^2m^2/t\epsilon^2} }\right)
\ee
 Equation (\ref{expecU}) together with the definition (\ref{mom})
gives the expectation value
\be
\langle \hat{P}_R^\epsilon\rangle =
\frac{\sin[P_R^0(x_k) \epsilon]}{\epsilon}\ e^{-t\epsilon^2/4}\ K(\epsilon,t,R^0).
\ee
This formula has the limits
\bes \lim_{t\rightarrow 0}\ \langle \hat{P}_R^\epsilon\rangle
&=& \sin(P_R^0(r_k)\epsilon)/\epsilon, \\
 \lim_{\epsilon\rightarrow 0}\ \langle \hat{P}_R^\epsilon\rangle
&=& P_R^0(r_k).
 \ees
The first shows that the semiclassical state in field space is
peaked at the corresponding phase space value. The second shows that
the field continuum limit of the momentum expectation value has the
appropriate peaked value in this state, even though only  field
translation operators exist in the representation we are using.

These semiclassical states defined at each point $r_i$ can now be used to give a state for the
entire radial lattice given any classical field configuration $R(r), P_R(r)$ by taking a
product of the point states over the lattice $\{r_k\}_{k=1}^N$, ie.
\be
|R(r),P(r)\rangle^{t,\epsilon} :=  \prod_{k=1}^N |R(r_k),P_R(r_k)\rangle^{t,\epsilon}_{r_k}
\ee
The simplest semiclassical state for both sets of fields is the product
\be
|\chi\rangle\equiv |R(r),P(r)\rangle^{t,\epsilon}\  |\phi(r),P_\phi(r)\rangle^{t,\delta}.
\label{scphys}
\ee
It is not difficult to configure entangled products of semiclassical states, given two distinct
classical solutions of the constraints.

Given that the expectation values in these states give the classical peaking values it follows that
the states peaked on  classical constraint free data, such as those constructed in Section II,
satisfy the expectation value
\be
\langle \chi | \hat{C}| \chi \rangle = 0 + {\cal O}(t^\sigma),
\ee
where $\sigma >1$.
We note also that the construction we have given is quite specific, and there
is control on the fluctuation $\langle  \hat{C}^2  \rangle $, which is itself a function of the width  $t$ of the point Gaussian states.  Its value can be tuned to  reduce the fluctuations.  It is in this sense that these states are approximately physical for sharply peaked states $t\ll 1$. One can now compute quantities such as the expectation value of the energy in such states, which is tedious  put possible.

\subsection{Effective equations}

 One approach for obtaining quantum gravity corrections is  to use semiclassical
 states   to derive ``effective constraints.''  These can then be used to  obtain  modified
 equations of motion in the usual way.  This approach is qualitatively related to the
 ideas underlying Ehrenfest's theorem in quantum mechanics.
One computes expectation values of the constraints in states peaked on classical configurations
that are not solutions of the initial value constraints to obtain
\be
\langle \phi,P_\phi, R,P_R |\hat{C} | \phi,P_\phi, R,P_R \rangle = C^{\text classical} (\phi,P_\phi, R,P_R)
+ {\cal O}(t^\sigma).
\ee
It is understood here that  the classical term on the right hand side is a function of the phase space configuration on which the states are peaked. The corrections are functions of the state's width $t$ in Planck units.

This  approach has been used for homogeneous isotropic cosmology \cite{effective-eqns}, but requires careful scrutiny in the field theory case here. One has to check that the Poisson algebra of the effective
constraints closes with the state width corrections included, since these corrections are
functions of the phase space peaking values of the semiclassical state. One of the goals  of the program presented here is to obtain consistent effective  equations of motion starting from semiclassical states.

\section{Summary and Discussion}

The paper contains a number of developments beyond earlier works on the spherically symmetric
gravity-scalar model:  (i) A construction of semiclassical states for the quantum theory of gravity coupled to a minimally coupled massless scalar field. These states may be utilized for constructing semiclassical  effective constraints, which may then be used for  classical numerical evolution. If used in conjunction
with constraint free data, such as that given in Sec. II.B, this construction gives the first known
examples of {\it physical} semiclassical states.  (ii) An alternative construction of the Hamiltonian operator in a fixed time gauge which can serve as a starting point for Monte Carlo simulations of the quantum theory. Since this model is a two-dimensional system
that is effectively written as a lattice theory, this method has the potential to reveal interesting non-perturbative quantum phenomena such as phase transitions. This is at present being studied.

The form of the Hamiltonian also reveals that matter-gravity entanglement is an inherent feature
of evolution in quantum gravity; the action of the Hamiltonian on a product state gives an entangled
state after a single evolution step.  In the full quantum problem the constraint operator forces a bipartite entanglement (with the two subsystems being the kinematical Hilbert spaces of matter and gravity degrees of freedom, respectively) already on the initial physical states. It is tempting to summarize this observation by modifying Wheeler's one line description of general relativity \cite{mtw}:  quantum gravity tells geometry and matter how to entangle.

The degree of entanglement may be computed in the usual way by  tracing a density matrix over either matter of geometry degrees of freedom. If a state describes a black hole  there is a second entropy that may be calculated, namely that obtained by tracing the (pure) density matrix over both geometry and matter degrees of freedom inside a  trapped region. This of course would be  different from the usual entanglement entropy where the trace in the interior applies only to the matter degrees of freedom.

It will be apparent to the reader that the extraction of physical results from this formalism  requires
further work. This could proceed along two distinct avenues. The first is the derivation of semiclassical
effective equations where one would compute expectation values of the constraint operators  in suitable
states to obtain ``effective'' constraints. These would then be used as the basis of a quantum gravity
corrected classical dynamics which could be integrated using numerical methods.  Among other things
it would be useful to see what becomes of the critical scaling observed  at the onset of black hole formation, especially the critical solution. Initial results using this approach \cite{vh-qgcrit, gz2} indicate that black holes form with a gap, but so far there are no results on the critical solution itself; with expected singularity avoidance, it is likely that this is replaced by a critical and unstable bound state -- a finely tuned boson star.

The second approach to studying dynamics  could utilize Monte Carlo methods.  The quantum model as formulated here is like a statistical mechanical system, with the difference that it is still a constrained theory.  One can imagine sampling from phase space such that the samples
are solutions of the constraints up to some threshold, along with the 
usual Monte Carlo selection criteria. Such an approach would have the potential to yield fundamental information concerning phase transitions.

\medskip

\noindent {\bf Acknowledgements} The work of V. H. was supported in part by the Natural Science and Engineering Research Council of Canada.

\end{document}